\documentclass {elsarticle}
\usepackage{graphicx}
\usepackage{color}
\usepackage{bm}
\usepackage{amsmath}
\usepackage{amssymb}
\usepackage{bm}
\usepackage{hyperref}
\usepackage{dcolumn}
\usepackage{slashed}

\begin{document}
\title{Introduction to Quantum-limited Parametric Amplification of Quantum Signals with Josephson Circuits}

\author[yu]{Michel Devoret}
\ead{michel.devoret@yale.edu}
\author[yu]{\text{and} Ananda Roy}
\ead{ananda.roy@yale.edu}
\address[yu]{Department of Applied Physics, Yale University, PO BOX 208284, New Haven, CT 06511}

\begin{abstract}
This short and opinionated review starts with a concept of quantum signals at microwave frequencies and focuses on the principle of linear parametric amplification. The amplification process arises from the dispersive nonlinearity of Josephson junctions driven with appropriate tones. We discuss two defining characteristics of these amplifiers: the number of modes receiving the signal, idler and pump waves and the number of independent ports through which these waves enter into the circuit. 
\end{abstract}

\maketitle

\section{Introduction}

Photons of microwave radiation in the band $3-12\ \rm{GHz}$ (25-100 mm wavelength)
have an  energy approximately 10$^{5}$ smaller than those of visible light.
Yet, at a temperature 2$\times $10$^{4}$ smaller than room temperature, now
routinely achievable with commercial dilution refrigerator, it is possible to
detect and process signals whose energy is equivalent to that of single
microwave photons \cite{Devoret_Schoelkopf_2013}. There are three advantages of single photon microwave
electronics when compared with quantum optics. First, signal envelopes with a relative bandwidth of few
percent at carrier frequencies of a few GHz can be controlled with much greater relative precision than their
equivalent at a few hundred of THz. This is because microwave generators
have better short term stability than lasers, and also because microwave
components are mechanically very stable, particularly when cooled, compared
with traditional optical components. Second, on-chip circuitry of single-photon microwave electronics can be well in the lumped
element regime and consequently, the control of spatial mode structure more easily
achieved than in the optical domain. Finally, there exists a simple, robust,
non-dissipative component, the Josephson tunnel junction, whose
non-linearity can dominate over the linear characteristics of the circuit at
the single photon level. Many quantum signal processing functions have
thus been realized, both digital and analog, and this short review will not
attempt to describe all of them. We will concentrate here on analog Josephson
amplifying devices pumped with one or several microwave tones. Only devices
that demonstrate linear amplification with added noise at the level of the
standard quantum limit \cite{Caves_1982} will be considered. These novel devices have taken
the work pioneered by B. Yurke at Bell labs 30 years ago \cite{Yurke_Simon_1988, Yurke_Whittaker_1989} to the point
where new original experiments can be performed successfully owing  to
Josephson amplifiers as the first link in the chain of measurements \cite{Vijay_Siddiqi_2011, Hatridge_Devoret_2013}. 

The article is organized as follows. Sec. \ref{sec_2} is devoted to the in-depth
description of signals that are considered quantum in the field of microwave
electronics. We describe in Sec. \ref{sec_3} the important theoretical tool of the Quantum Langevin Equation with which the amplifier characteristics can be calculated, starting from the circuit Hamiltonian and the coupling parameters of its ports. 
Then, in the following Sec. \ref{sec_4}, we will introduce the notion
of effective parametric amplifier Hamiltonian. This will lead us to discuss
the important distinction between the degenerate and non-degenerate
amplifiers that arises from a fundamental difference in the in the degrees
of freedom of the two devices. The practical implementation of amplifiers
will be treated in Sec. \ref{sec_5}. Finally, in Sec. \ref{sec_6}, after a concluding summary, we
will indicate the perspectives of the field.

\section{Quantum electromagnetic signals propagating along transmission lines}
\label{sec_2}

\begin{figure}
\centering
\includegraphics[width = 0.9\textwidth]{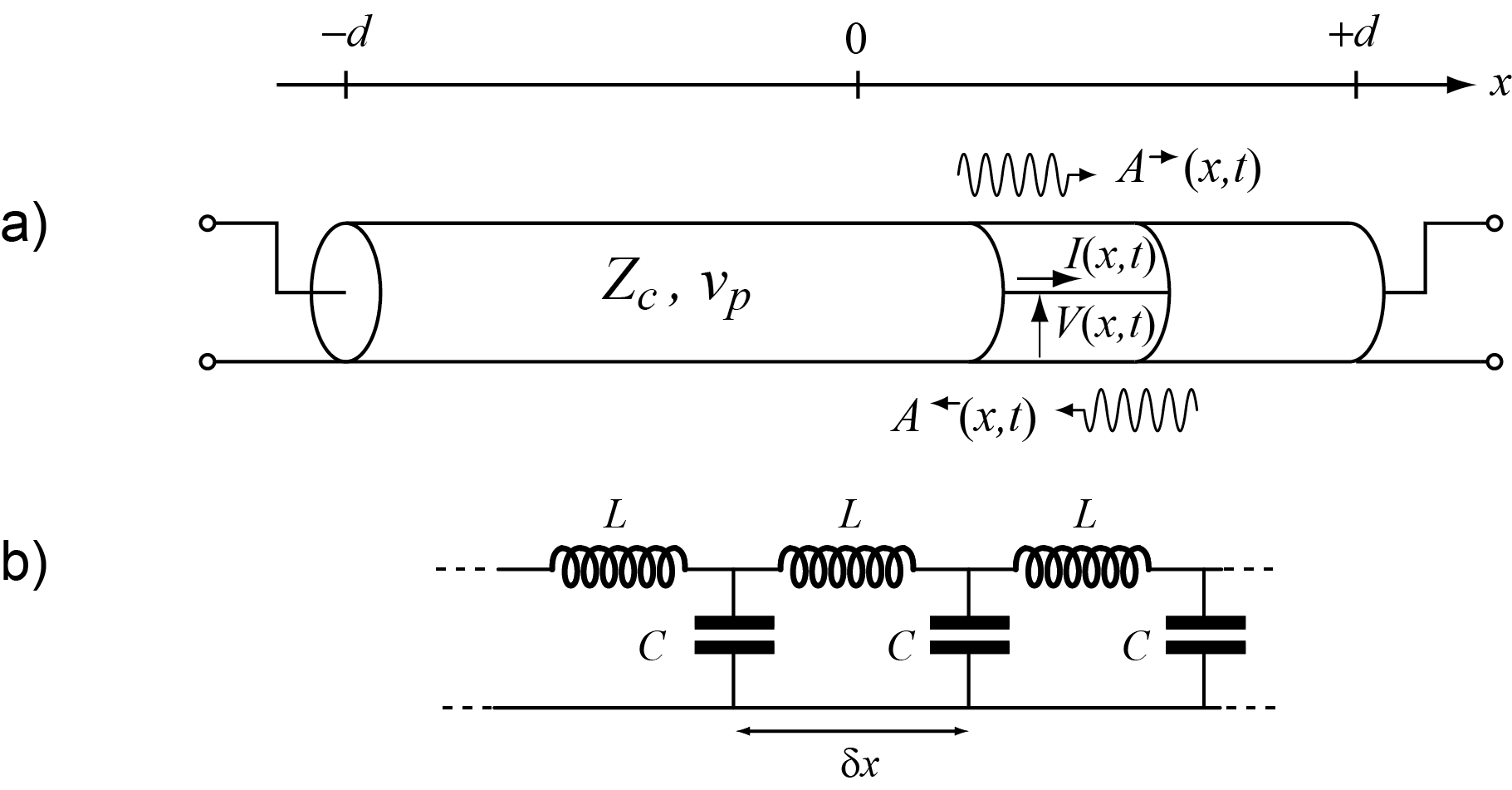}
\caption{\label{Fig_1} (a) Electromagnetic transmission line implemented as a
coaxial cable. The parameter $x$ denotes the position along the line, $I$
denotes the current along the line in the positive direction and $V$ the
voltage between the inner and outer conductors. The characteristic impedance
and the propagation velocity are denoted by $Z_{c}$ and $v_{p}$,
respectively. The line has a continuous density of modes in the limit where
its length $2d\rightarrow \infty $. In (b), a ladder circuit model with cell
dimension $\protect\delta x$ models the infinite transmission line. Its
capacitance and inductance per unit length are given by $L_{\ell }=L/\protect%
\delta x$ and $C_{\ell }=C/\protect\delta x$, respectively. In the limit
where the signal frequency $\protect\omega $ is small compared to $1/\protect%
\sqrt{LC}$, $Z_{c}=\protect\sqrt{L/C}$ and $v_{p}=1/\protect\sqrt{L_{\ell
}C_{\ell }}$. }
\end{figure}

Crudely speaking, quantum signals are electromagnetic excitations of a
transmission line that involve only a few photons. The  state of these
excitations must display some degree of quantum purity for the signals to
carry quantum information, which is the subject of interest in Josephson
circuits. In this section, we provide the basic mathematical background for
the concept of photon applied to microwave electromagnetic excitations \cite{Gardiner_Zoller_2004, Wall_Milburn_2008}. We
start by considering an infinite transmission line, a one-dimensional
electromagnetic medium characterized by a propagation velocity $v_{p}$ and a
characteristic impedance $Z_{c}$. A microwave coaxial line serves as the
canonical example of such medium (see Fig. \ref{Fig_1}). 

Position along the line is indexed by the real number $x\in ( -\infty
,+\infty )$. We suppose that the line is ideal, with both $v_{p}$ and 
$Z_{c}$ independent of frequency $\omega $. It is convenient to combine the
voltage $V\left( x,t\right) $ across the line with the current $I\left(
x,t\right) $ along the line in the $\ +x$ direction into the so-called
propagating wave amplitude

\begin{equation}
A^{\rightleftarrows }\left( x,t\right) =\frac{1}{2}\left[ \frac{1}{\sqrt{%
Z_{c}}}V\left( x,t\right) \pm \sqrt{Z_{c}}I\left( x,t\right) \right].
\end{equation}%
Here, the superscript arrows refer to the two directions of propagation
along the line. The wave amplitude, whose dimension is [watt]$^{1/2}$, is
such that its square is the energy flux of waves traveling in the direction
indicated by the arrow. This interpretation is justified by the formula
giving the total energy flux at time $t$ crossing the location $x$, which is
the equivalent of the Poynting vector for a one-dimensional medium:

\begin{equation}
P\left( x,t\right) =\left\vert A^{\rightarrow }\left( x,t\right) \right\vert
^{2}-\left\vert A^{\leftarrow }\left( x,t\right) \right\vert ^{2}=V\left(
x,t\right) I\left( x,t\right).
\end{equation}

Because the voltage and current obey the equivalent of Maxwell's equations%
\begin{eqnarray}
\frac{\partial V}{\partial x} &=&-L_{\ell }\frac{\partial I}{\partial t}, \\
\frac{\partial I}{\partial x} &=&-C_{\ell }\frac{\partial V}{\partial t},
\end{eqnarray}%
where $\left( L_{\ell }C_{\ell }\right) ^{-1/2}=v_{p}$ and $\left( L_{\ell
}/C_{\ell }\right) ^{1/2}=Z_{c}$, the wave amplitude has the space-time
translation invariance properties of traveling electromagnetic waves

\begin{equation}
A^{\rightleftarrows }\left( x,t\right) =A^{\rightleftarrows }\left( 0,t\mp
x/v_{p}\right) =A^{\rightleftarrows }\left( x\mp v_{p}t,0\right).
\end{equation}%
This shows that the entire field in the line is completely described by the
left-moving and right-moving components at time $t=0$ for all position $x$.
Equivalently, the field is described by the left-moving and right-moving
components at $x=0$ for all times $t$. We now introduce the wave amplitude
operator, which describes the electromagnetic field of the line
quantum-mechanically

\begin{equation}
A^{\rightleftarrows }\left( x,t\right) \rightarrow \mathbf{A}^{\rightleftarrows
}\left( x\right).
\end{equation}

Instead of a space index, we can give the wave amplitude operator a time
index, which is trivially deduced from the space index

\begin{equation}
\mathbf{A}^{\rightleftarrows }\left( x\right) \Longleftrightarrow \mathbf{A}%
^{\rightleftarrows }\left( t=\pm \frac{x}{v_{p}}\right).
\end{equation}%
The commutation properties of the wave amplitude operator, which is
hermitian, are rather complex. The route to the mathematical introduction of
the microwave photon is easier if we first make a detour through Fourier
space. We define

\begin{equation}
\mathbf{A}^{\rightleftarrows }\left[ \omega \right] =\frac{1}{\sqrt{2\pi }}%
\int\nolimits_{-\infty }^{+\infty }dt\ e^{i\omega t}\mathbf{A}^{\rightleftarrows
}\left( t\right),
\end{equation}%
which describes the wave amplitude operator in the frequency domain (no real
time dynamics is considered here, as the time index is equivalent to a space
index and the reciprocal space equivalent to the wave vector space). We now
introduce field ladder operators

\begin{equation}
\mathbf{a}^{l}\left[ \omega \right] =\frac{1}{\sqrt{\hbar \left\vert \omega
\right\vert /2}}\mathbf{A}^{l}\left[ \omega \right],
\end{equation}%
where the superscripts $l=\pm 1$ denote the direction of propagation $%
\rightarrow $ or $\leftarrow $ (we will generalize later the index $l$ to include the
line number, i.e. spatial mode number). The field ladder operators $\mathbf{a}^{l}%
\left[ \omega \right] $ have commutation relations bearing a marked
resemblance to the ladder operators of a set of standing wave harmonic
oscillators

\begin{equation}
\left[ \mathbf{a}^{l_{1}}\left[ \omega _{1}\right] ,\mathbf{a}^{l_{2}}\left[ \omega _{2}\right]
\right] =\mathrm{sgn}\left( \omega _{1}-\omega _{2}\right) \delta \left(
\omega _{1}+\omega _{2}\right) \delta _{l_{1},l_{2}}.
\end{equation}%
This is somewhat clearer when we take into account that 
\begin{equation}
\mathbf{a}^{l}\left[ \omega \right] ^{\dag }=\mathbf{a}^{l}\left[ -\omega \right] .
\end{equation}%
Yet, these operators do not correspond directly to the traveling photon
ladder operators. We go back in the time domain and introduce%
\begin{eqnarray}
\mathbf{a}^{l}\left( t\right) &=&\frac{1}{\sqrt{2\pi }}\int\nolimits_{-\infty
}^{+\infty }d\omega e^{-i\omega t}\mathbf{a}^l\left[ \omega \right] =\tilde{\mathbf{a}}
^{l}\left( t\right) +\tilde{\mathbf{a}}^{l}\left( t\right) ^{\dag },\label{alpropagating} \\
\tilde{\mathbf{a}}^{l}\left( t\right)&=&\frac{1}{\sqrt{2\pi }}\int\nolimits_{0}^{+
\infty }d\omega e^{-i\omega t}\mathbf{a}^l\left[ \omega \right],\\
\tilde{\mathbf{a}}^{l}\left( t\right) ^{\dag } &=&\frac{1}{\sqrt{2\pi }}
\int\nolimits_{-\infty }^{0}d\omega e^{-i\omega t}\mathbf{a}^l\left[ \omega \right].
\end{eqnarray}

It is important to note that with our notations $\tilde{\mathbf{a}}^{l}\left(
t\right) $ is NOT the inverse Fourier transform of $\mathbf{a}^{l}\left[ \omega %
\right] $, since it involves only positive frequencies $\omega $. The
inverse Fourier transform of $\mathbf{a}^{l}\left[ \omega \right] $ is $\mathbf{a}^{l}\left(
t\right) $, which is a \emph{hermitian} operator. The dimension of the
operators $\tilde{\mathbf{a}}^{l}\left( t\right) $ is the inverse square root of time
and it would be tempting to interpret them as photon flux amplitudes.
However, in the time domain, the commutation relations of the field ladder
operators do not take the usual bosonic form of a scalar field

\begin{eqnarray}
\left[ \tilde{\mathbf{a}}^{l}\left( t_{1}\right) ,\tilde{\mathbf{a}}^{l}\left( t_{2}\right)
^{\dag }\right] &=&\frac{1}{2\pi }\left[ \int\nolimits_{0}^{+\infty }d\omega
_{1}e^{-i\omega _{1}t_{1}}\mathbf{a}\left[ \omega _{1}\right] ,\int\nolimits_{-\infty
}^{0}d\omega _{2}e^{-i\omega _{2}t_{2}}\mathbf{a}\left[ \omega _{2}\right] \right]\nonumber\\
&=&\frac{1}{2}\delta \left( t_{1}-t_{2}\right) +\frac{i}{2\pi }\text{\textrm{%
p.p.}}\left( \frac{1}{t_{1}-t_{2}}\right).
\end{eqnarray}%
On the other hand,%
\begin{equation}
\left[ \mathbf{a}^{l_{1}}\left( t_{1}\right) ,\mathbf{a}^{l_{2}}\left( t_{2}\right) \right]
=\frac{i}{\pi} {\rm{p.p}}\frac{1}{t_1-t_2}\delta _{l_{1},l_{2}}.
\end{equation}%
In order to properly define the photons of the line, one needs to introduce
an orthonormal signal basis consisting of \textquotedblleft
first-quantization\textquotedblright\ wavelets $w_{mp}^{l}\left( t\right) $
such that%
\begin{eqnarray}
\int\nolimits_{-\infty }^{+\infty }dt\;w_{m_{1}p_{1}}^{l_{1}}\left( t\right)
w_{m_{2}p_{2}}^{l_{2}}\left( t\right) ^{\ast } &=&\delta
_{m_{1},m_{2}}\delta _{p_{1},p_{2}}\delta _{l_{1},l_{2}}, \\
w_{mp}^{l}\left( t\right) ^{\ast } &=&w_{-mp}^{l}\left( t\right) , \\
\sum_{m=-\infty }^{+\infty }\sum_{p=-\infty }^{+\infty }w_{mp}^{l}\left(
t_{1}\right) w_{-mp}^{l}\left( t_{2}\right) &=&\delta \left(
t_{1}-t_{2}\right).
\end{eqnarray}%
The pair of indices $\left( \left\vert
m\right\vert ,p\right) \in \mathbb{N}^{+}\times \mathbb{Z}$ defines a
propagating temporal mode of the line, and the combined amplitudes of the
two corresponding wavelets can be seen as an elementary degree of freedom of
the field. There are two conjugate wavelets per mode since the phase space
of each mode is bi-dimensional.

It is necessary to request that the support of $%
w_{mp}^{l}\left[ \omega \right] $, the Fourier transform of $%
w_{mp}^{l}\left( t\right) $, is entirely contained in the positive frequency
sector if $m>0$ and in the negative frequency sector if $m<0$.%
\begin{eqnarray}
w_{mp}^{l}\left[ \omega \right] &=&w_{mp}^{l}\left[ \omega \right] \Theta
\left( \omega \right) \qquad \textrm{if}\qquad m>0, \\
w_{mp}^{l}\left[ \omega \right] &=&w_{mp}^{l}\left[ \omega \right] \Theta
\left( -\omega \right) \qquad \textrm{if}\qquad m<0.
\end{eqnarray}%
In these last expressions, $\Theta \left( \omega \right) $ is the Heaviside
function \footnote{The index value $m=0$ corresponds to special wavelets that have to be treated separately.}.

This complete wavelet basis is a purely classical signal processing concept
and its existence solely results from the property of the signals to be
square-integrable functions. Any continuous signal $f\left( t\right) $ such
that $\int_{-\infty }^{+\infty }\left\vert f\left( t\right) \right\vert
^{2}dt<\infty$
can indeed be decomposed into a countable infinite number of elementary
signals 
\begin{eqnarray}
f\left( t\right) &=&\sum_{m=-\infty }^{+\infty }\sum_{p=-\infty }^{+\infty
}f_{-mp}w_{mp}\left( t\right), \\
f_{mp} &=&\int\nolimits_{-\infty }^{+\infty }dt\;w_{mp}\left( t\right)
f\left( t\right).\label{fmus}
\end{eqnarray}%
A common example of such a wavelet is the Shannon wavelet

\begin{equation}
\mathfrak{w}_{mp}\left( t\right) =2\sqrt{\frac{\tau }{2\pi }}\frac{\sin
\left( \frac{\pi }{\tau }(t-p\tau )\right)}{t}e^{i2\pi mt/\tau }
\end{equation}%
whose Fourier transform is 
\begin{equation}
\mathfrak{w}_{mp}\left[ \omega \right] =\sqrt{\frac{\tau }{2\pi }}1_{\frac{%
2\pi }{\tau }(m-1/2),\frac{2\pi }{\tau }(m+1/2)}\left( \omega \right)
e^{ip\omega \tau },
\end{equation}%
where $1_{x_{1},x_{2}}\left( x\right) $ is the indicator function which is $%
0 $ everywhere except in the interval $\left[ x_{1},x_{2}\right] $, where
its value is unity. Many other useful bases, involving more continuous
wavelets, exist \cite{Mallat_1999}. In the above example, the center frequency
and time location of the wavelet is $2\pi m/\tau $ and $p\tau $, respectively
(in order to form a complete basis, the pitch in frequency $\Delta \omega $
and pitch in time $\Delta t$ of the wavelet basis has to satisfy $\Delta
\omega .\Delta t\leq 2\pi $) . 

The discreteness of the signal component indices is the justification for
the term \textquotedblleft first-quantization\textquotedblright\ and no
quantum mechanics is involved here since all functions are at this stage
c-number valued. Second-quantization intervenes when we define the discrete ladder
field operators, with indices $m>0$ and $p$
\begin{eqnarray}
\bm{\psi}_{mp}^{l} &=&= \int\nolimits_{-\infty }^{+\infty }d\omega
w_{mp}^{l}\left( \omega \right) \mathbf{a}^{l}\left( \omega \right), \\
\bm{\psi}_{-mp}^{l} &=&\bm{\psi}_{mp}^{l\dag }.
\end{eqnarray}
We introduce the short-hand $\mu = (l, |m|, p)$ as the index of the spatio-temporal mode, also called the flying oscillator. The photon-number operator is given by:  
\begin{equation}
\mathbf{n}_{\mu }=\bm{\psi}_{\mu }^{\dag }\bm{\psi}_{\mu }
\end{equation}
and the discrete ladder operators $\bm{\psi}_{\mu}$ satisfy the same commutation relation as
standing mode ladder operators:
\begin{eqnarray}
\left[ \bm{\psi}_{\mu_1},\bm{\psi}_{\mu_2}^\dag %
\right] &=&\int\nolimits_{-\infty }^{+\infty }\int\nolimits_{-\infty
}^{+\infty }d\omega_{1}d\omega_{2}w_{m_{1}p_{1}}^{l_{1}}\left( \omega_{1}\right)
w_{m_{2}p_{2}}^{l_{2}}\left( \omega_{2}\right) ^{* }[\mathbf{a}^{l_{1}}\left(
\omega_{1}\right), {\mathbf{a}^{l_{2}}\left( \omega_{2}\right)}\nonumber] \\
&=&\delta _{\mu_1, \mu_2}.
\end{eqnarray}%
An important
remark can be made: if the photon amplitude operator $\bm{\psi}_{\mu }$ is
non-hermitian, this is only because its first quantization component $%
w_{mp}^{l}\left( t \right) $ is complex. Its second quantization
component $\mathbf{a}^{l}\left( t \right) $ is an hermitian operator. It is also
important to note that, in general, the frequency of a photon is
ill-defined, in contrast with what could be inferred from elementary
introductions to quantum mechanics. This feature happens as soon as the
duration of the wavelet corresponding to that particular photon is not very
long compared with the inverse of the wavelet center frequency. Thus the
concept of photon for a propagating signal has to be clearly distinguished
from an energy quantum. A propagating photon is an elementary excitation of
the field carrying a quantum of action, and corresponds to a field
wavefunction orthogonal to the vacuum.%
\begin{eqnarray}
\left\vert \Psi _{1\mu }\right\rangle &=&\bm{\psi}_{\mu }^{\dag }\left\vert 
\mathrm{vac}\right\rangle, \\
\left\langle \mathrm{vac}|\Psi _{1\mu }\right\rangle &=&0.
\end{eqnarray}%
A wavelet can contain several photons in mode $\mu $ 
\begin{equation}
\left\vert \Psi _{n\mu }\right\rangle =\frac{1}{\sqrt{n!}}\left( \bm{\psi}%
_{\mu }^{\dag }\right) ^{n}\left\vert \mathrm{vac}\right\rangle
\end{equation}%
and each multi-photon state (Fock state) is orthogonal to the others\newline
\begin{equation}
\left\langle \Psi _{n_{2}\mu }|\Psi _{n_{1}\mu }\right\rangle =\delta
_{n_{1}n_{2}}.
\end{equation}%
Several modes can simultaneously be excited%
\begin{equation}
\left\vert \Psi _{n_1,\mu _{1};n_2,\mu _{2};n_3,\mu _{3};\ldots}\right\rangle =\frac{1}{%
\sqrt{n_{1}!}}\left( \bm{\psi}_{\mu _{1}}^{\dag }\right) ^{n_{1}}\frac{1}{%
\sqrt{n_{2}!}}\left( \bm{\psi}_{\mu _{2}}^{\dag }\right) ^{n_{2}}\frac{1}{%
\sqrt{n_{3}!}}\left( \bm{\psi}_{\mu _{3}}^{\dag }\right)
^{n_{3}}\ldots\left\vert \mathrm{vac}\right\rangle.
\end{equation}%
The sequence of indices $\sigma =\left( n_{1},\mu _{1};n_{2},\mu _{2};n_{3},\mu
_{3};\dots\right) $ is a mode photon occupancy configuration. Finally, the
most general wavefunction of the field of the transmission line(s) is a
superposition of all field photon configurations in all the spatio-temporal
modes of the line(s):%
\begin{equation}
\left\vert \Psi \right\rangle =\sum_{\sigma }C_{\sigma }\left\vert \Psi
_{\sigma }\right\rangle.
\end{equation}%
There are exponentially many more quantum coefficients $C_{\sigma }$ than
the classical coefficients $f_{\mu }$ in Eq.~(\ref{fmus})! And it is also important to
understand that a state with a well defined number of photons in a certain
wavelet basis can be fully entangled in another basis.

A wavelet can also support a so-called coherent state instead of a well
defined number of photons:

\begin{eqnarray}
\left\vert \alpha _{\mu }\right\rangle &=&e^{-\left\vert \alpha _{\mu
}\right\vert ^{2}/2}\sum_{n}\frac{\alpha _{\mu }^{n/2}}{\sqrt{n!}}\left\vert
\Psi _{n\mu }\right\rangle, \\
&=&e^{-\left\vert \alpha _{\mu }\right\vert ^{2}/2}e^{\alpha _{\mu }\bm{\psi}
_{\mu }^{\dag }}\left\vert \mathrm{vac}\right\rangle,
\end{eqnarray}%
and if all wavelets are in a coherent state, we obtain a coherent field state%
\begin{eqnarray}
\left\vert \Psi _{\left\{ \alpha \right\} }\right\rangle
&=&\prod\limits_{\mu }\left\vert \alpha _{\mu }\right\rangle \\
&=&e^{-\sum_{\mu }\left( \left\vert \alpha _{\mu }\right\vert ^{2}/2-\alpha
_{\mu }\bm{\psi}%
_{\mu }^{\dag }\right) }\left\vert \mathrm{vac}\right\rangle.
\label{General_coherent_state}
\end{eqnarray}%
Thus, the set of complex coefficients $\alpha _{\mu }$ plays the role of the
coefficients $f_{\mu }$ in Eq.~(\ref{fmus}). Somewhat surprisingly, this property of being a
coherent state remains true in \emph{every} wavelet basis (as can be
inferred from the quadratic form in the exponent of Eq. [\ref%
{General_coherent_state}]). 

The state of the line is in general not pure and must be described by a
density matrix $\rho _{\sigma \sigma ^{\prime }}$. This ultimate quantum
field description tool leads to the important notion of information
contained in the signal.

In general, in quantum mechanics, we can define for a system with a
finite-dimension Hilbert space, the Shannon$-$Von Neumann entropy%
\begin{equation}
S=-\mathrm{tr}\rho \ln \rho. 
\end{equation}%
The information contained in the system is then straightforwardly computed as%
\begin{equation}
\mathcal{I}=S\left( \rho _{\mathrm{mix}}\right) -S\left( \rho \right),
\end{equation}%
where $\rho _{\mathrm{mix}}$ is the fully mixed state in which all basis
states are equiprobable, with no off-diagonal correlations. The extension of
these ideas to a transmission line on which a signal propagates is not
trivial since the number of temporal modes is infinite and each temporal
mode has a Hilbert space with infinite dimensionality. Some constraints need
to be provided, for instance a fixed total energy for both $\rho $ and $\rho
_{\mathrm{mix}}$. We can also, in another instance, fix the maximum number
of excitation in each temporal mode. Supposing that the maximum number of
excitations is unity in the domain $(\left\vert m\right\vert ,p)\in \left\{
1,2,..,M\right\} \otimes \left\{ -P,...,+P\right\} $ and that other modes
are in the vacuum state, then, for a state of the line characterized by an
average photon number $\left\langle n_{\left\vert m\right\vert
p}\right\rangle $ per mode%
\begin{equation}
\mathcal{I}=\sum_{m=1}^{M}\sum_{p=-P}^{P}\mathcal{I}_{b}\left( \left\langle
n_{\left\vert m\right\vert p}\right\rangle -1/2\right),
\end{equation}%
where $\mathcal{I}_{b}\left( \left\langle X\right\rangle \right) $ the
information contained in a stochastic binary variable $X=\pm 1$:%
\begin{equation}
\mathcal{I}_{b}\left( x\right) =\log _{2}\left[ \sqrt{1-x^{2}}\left( \frac{%
1+\left\vert x\right\vert }{1-\left\vert x\right\vert }\right) ^{\frac{%
\left\vert x\right\vert }{2}}\right] .
\end{equation}
We refer the reader to \cite{Holevo_2013} for a more complete description of the information carried by quantum signals. 

\section{Quantum Langevin Equation}
\label{sec_3}
\begin{figure}
\centering
\includegraphics[width = 0.7\textwidth]{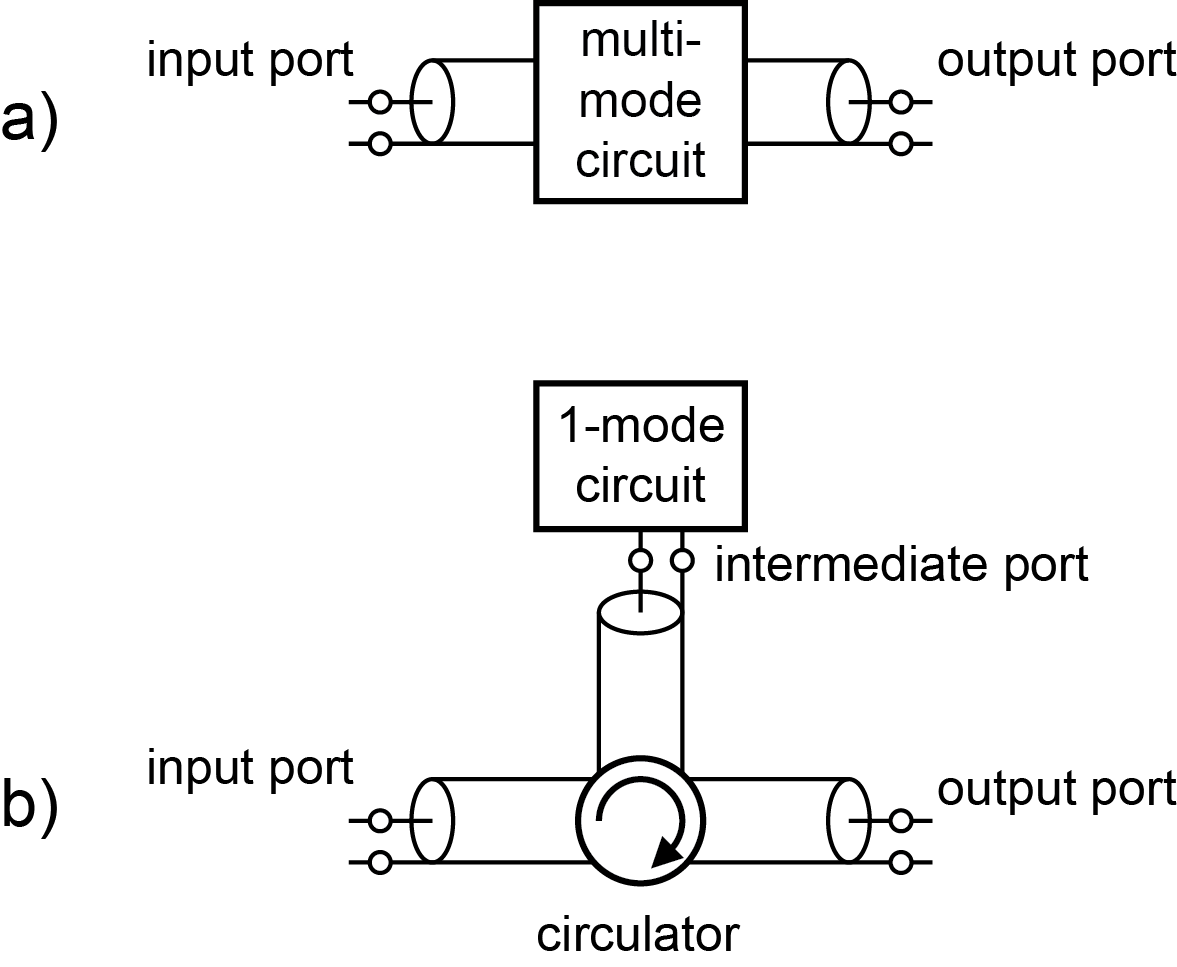}
\caption{\label{Fig_2}	Circuit modes and ports: a) We are interested
in quantum signal processing circuits in lumped element regime, which
possess in general several standing modes. Input lines and output lines are
attached to ports. In the simplest case b), a one-mode circuit communicates
with the outside through only one port. An ideal circulator separates the
input from the output.}


\end{figure}

The main role of the previous section was to introduce the concept of
quantum electromagnetic fields propagating along a transmission line at
microwave frequencies. The elementary excitations of these fields, microwave
photons, can be seen as the carriers of the information transmitted by the
propagating field. An amplifier is a particular case of a signal processing
device. In general, a signal processing device is a lumped element circuit
that is connected to two semi-infinite transmission lines [see Fig. \ref{Fig_2} (a)]. It
receives from the input line the propagating signal carrying the information
to be processed and re-emits in the output line another signal carrying the
result of the information processing. A crucial ingredient, in the
description of the mapping of the input signal into the output signal, is
the coupling between the circuit, which houses standing electromagnetic
modes, and the transmission lines, which support propagating modes. This
coupling is dealt with theoretically through the Quantum Langevin Equation.
We will first consider its simplest version in which a lumped element
circuit with only one electromagnetic mode is coupled to only one
semi-infinite transmission line [Fig. \ref{Fig_2} (b)].

In this so-called
one-mode, one-port configuration, the signal processing occurs as a
transformation of the incoming wave into the reflected outgoing wave.
However, in order to utilize the action of the one-mode, one-port circuit, a
non-reciprocal linear device called a circulator has to be added in order to
separate the incoming and outgoing waves into two independent transmission
lines [Fig. \ref{Fig_2} (b)]. This circuit configuration also provides a way to model directional, through amplifiers \cite{Courty_Reynaud_1999}. 
We can leave the modeling of the
circulator aside for the moment (it is a 3-port device), and just write the
equation linking the quantum amplitude $\mathbf{a}$ of the circuit, satisfying $\left[
\mathbf{a},\mathbf{a}^{\dag }\right] =1$, to the incoming quantum field amplitude of the line $%
\mathbf{a}^{\rm{in}}\left( t\right) $ at the unique port (this is the generic name of the
interface between a semi-infinite transmission line and a lumped element
circuit)

\begin{equation}
\frac{d\mathbf{a}}{dt}\underset{%
\begin{array}{c}
\text{{\tiny Markov}} \\ 
\text{{\tiny RWA}}%
\end{array}%
}{=}\frac{i}{\hbar }\left[ \mathbf{H},\mathbf{a}\right] -\frac{\kappa }{2}\mathbf{a}+\sqrt{\kappa }%
\mathbf{a}^{\rm{in}}\left( t\right) ,  \label{Quantum_Langevin}
\end{equation}%
\newline
where $\mathbf{a}^{\rm{in}}$ is the traveling photon amplitude defined in the last
section (Eq.~(\ref{alpropagating})), the superscript ``in'' referring to the sense of propagation coming
into the circuit. The equation (\ref{Quantum_Langevin}) is a special case of
the Quantum Langevin Equation (QLE) \cite{Gardiner_Zoller_2004} and its right handside has three terms.
The first term corresponds to that of the usual Heisenberg equation of
motion for an operator in quantum mechanics, in which $\mathbf{H}$ is the Hamiltonian
of the circuit, written as a function of the conjugate ladder operators $\mathbf{a}$
and $\mathbf{a}^{\dag }$. The circuit can be arbitrarily non-linear and thus $\mathbf{H}$ will
be in general more sophisticated than the simple quadratic term $\hbar
\omega _{a}\mathbf{a}^{\dag }\mathbf{a}$ of the quantum harmonic oscillator. Next, the second
term is a damping term specific to both the open nature of the system, and
the linear coupling between the circuit and the transmission line. Its
remarkable simple form requires two assumptions: i) The so-called Markov
approximation which considers that the coupling of the system with the
environment is ``ohmic": the density of modes of the environment can be
considered white across the set of circuit transition frequencies, as in an ideal resistance. ii) The
coupling is also supposed to be weak in the sense that $\kappa $ is much
smaller than any transition frequency between the energy levels of the
lumped circuit. This is the so-called Rotating Wave Approximation (RWA).
Finally, the third and last term on the right handside of the Quantum
Langevin Equation describes the role of the incoming field as a drive for
the circuit. It represents the counterpart of the energy loss modeled by the
second term. Although approximations are made, the equation respects the
important commutation relation of the ladder operators:
\begin{equation}
\left[ \mathbf{a}\left( t\right) ,\mathbf{a}\left( t\right) ^{\dag }\right] =1
\end{equation}%
at all times $t$.

The incoming driving field has in general three components which are treated
on equal footing by the QLE: i) the deterministic signal to be processed,
ii) thermal or parasitic noise accompanying the information-carrying signal,
and iii) quantum noise, or, in other words, the zero-point fluctuations of
the field of the semi-infinite transmission line. The inclusion of this last
component is implemented implicitly in that the Quantum Langevin Equation is
an operator equation, in contrast with the Classical Langevin Equation which
is just a differential equation for a c-number function, albeit stochastic.
Note that the coefficient $\sqrt{\kappa }$ in front of the propagating field
amplitude embodies single-handedly the fluctuation-dissipation theorem: the
rate at which energy is radiated away from the circuit (the coefficient $%
\kappa $ of the second term) has to be tightly linked to the coupling
constant with which random radiation -- emitted from the black body that the
line plays the role of -- corrupt the purity of the state of the circuit. If one wonders
why $\kappa $ appears under a square root in this coupling coefficient,
one just needs to remember that while $\mathbf{a}$ is a dimensionless standing photon
number amplitude, $\mathbf{a}^{\rm{in}}\left( t\right) $ is the dimensioned amplitude
corresponding to a photon flux. Consequently, when the semi-infinite line is
in thermal equilibrium (input signal is only black-body noise with
temperature $T$), the following relation involving the anticommutator $%
\left\{ \square ,\square \right\} $ holds%
\begin{equation}
\left\langle \left\{ \mathbf{a}^{\rm{in}}[\omega _{1}],\mathbf{a}^{\rm{in}}[\omega _{2}]\right\}
\right\rangle _{T}\,=\coth \frac{\hbar \left(| \omega _{1}-\omega _{2}|\right) 
}{4k_{B}T}\delta \left( \omega _{1}+\omega _{2}\right),
\end{equation}%
where $k_{B}$ is Boltzmann constant, $\left\{ \mathbf{A},\mathbf{B}\right\} =\mathbf{AB}+\mathbf{BA}$ and $%
\left\langle ...\right\rangle _{T}$ the average in the thermal state. Given
an operating temperature around $20~\mathrm{mK}$, this expression shows that
the quantum fluctuations become fully dominant over thermal fluctuations at
frequencies above a GHz.

In principle, one can integrate the QLE and express the circuit variable $%
\mathbf{a}(t)$ in terms of the incoming field $\mathbf{a}^{\rm{in}}\left( t\right) $. The output
field can be obtained from the Input-Output Equation%
\begin{equation}
\label{input-output}
\sqrt{\kappa }\mathbf{a}=\mathbf{a}^{\rm{in}}\left( t\right) +\mathbf{a}^{\rm{out}}\left( t\right).
\end{equation}%
Again, the appearance of the simple coefficient $\sqrt{\kappa }$ in this
relation results from the Markov approximation.

When the circuit is a simple harmonic oscillator $H=\hbar \omega _{0}\mathbf{a}^{\dag
}\mathbf{a}$, see Fig. 3, the elimination of $\mathbf{a}$ between input and output can be
carried out fully at the analytical level and one obtains%
\begin{equation}
\left( \frac{d}{dt}+i\omega _{0}+\kappa /2\right) \mathbf{a}^{\rm{out}}\left( t\right)
=-\left( \frac{d}{dt}+i\omega _{0}-\kappa /2\right) \mathbf{a}^{\rm{in}}\left( t\right).
\end{equation}%
Going to the Fourier domain, one obtains the reflection coefficient $r\left(
\omega \right) $

\begin{eqnarray}
\mathbf{a}^{\rm{out}}\left[ \omega \right]  &=&r\left( \omega \right) \mathbf{a}^{\rm{in}}\left[ \omega %
\right], \\
r_{\mathrm{RWA}}\left( \omega \right)  &=&-\frac{\omega -\omega _{0}-i\kappa
/2}{\omega -\omega _{0}+i\kappa /2}.
\end{eqnarray}

\begin{figure}
\centering
\includegraphics[width = 0.4\textwidth]{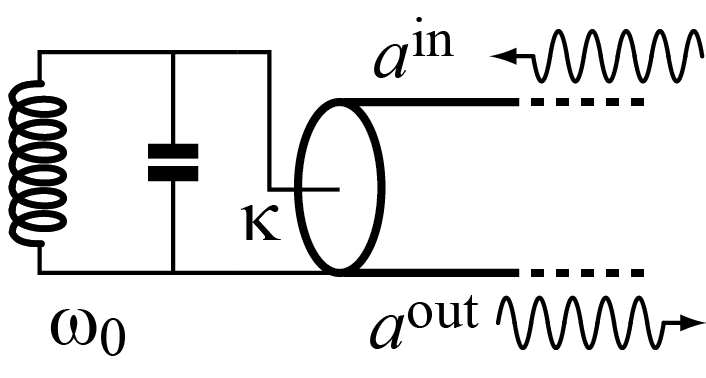}
\caption{\label{Fig_3} Schematic of the circuit corresponding to a 1-port 1-mode passive linear device.}
\end{figure}

The causality property of the circuit, which expresses the fact that it
cannot produce a response \textit{before} being submitted to a stimulus, is
implemented here by the analytic property of the complex function $r_{%
\mathrm{RWA}}(\omega )$: its pole is in the lower half complex plane while
its zero is in the upper half. On the other hand, the fact that the
reflection coefficient has only one pole instead of a pair is an artefact of
RWA. As a matter of fact, when the circuit is linear as in Fig. \ref{Fig_3}, one can
compute exactly the reflection coefficient using a more elaborate form of
QLE without RWA, while keeping the Markov approximation. One then obtains
the expression possessing the necessary pair of poles with values $\omega
_{\pm }=\left( -i\kappa \pm \sqrt{-\kappa ^{2}-4\omega _{0}^{2}}\right) /2$:

\begin{equation}
r\left( \omega \right) =-\frac{\omega ^{2}-\omega _{0}^{2}-i\kappa \omega }{%
\omega ^{2}-\omega _{0}^{2}+i\kappa \omega }.
\end{equation}%
It is easy to see that in this last equation, $r\left( \omega \right) $
reduces to the single pole expression $r_{\mathrm{RWA}}\left( \omega \right) 
$ when $\omega $ is such that $\left\vert 1-\omega /\omega _{0}\right\vert
\ll 1$ and in the underdamped limit $\kappa /\omega _{0}\ll 1$.

Continuing to work in the framework of both RWA and the Markov
approximation, one can easily deal with more than one circuit mode and more
than one semi-infinite line. Denoting by $M$ the circuit standing mode index
and $P$ the port index, one obtains the multi-mode, multi-port generalized
QLE:%
\begin{equation}
\frac{d}{dt}\mathbf{a}_{M}=\frac{i}{\hbar }\left[ \mathbf{H},\mathbf{a}_{M}\right] +\sum_{P}\left[ -%
\frac{\kappa _{MP}}{2}\mathbf{a}_{M}+\varepsilon _{MP}\sqrt{\kappa _{MP}}%
\mathbf{a}_{P}^{\rm{in}}\left( t\right) \right].
\end{equation}

\begin{figure}
\centering
\includegraphics[width = 0.7\textwidth]{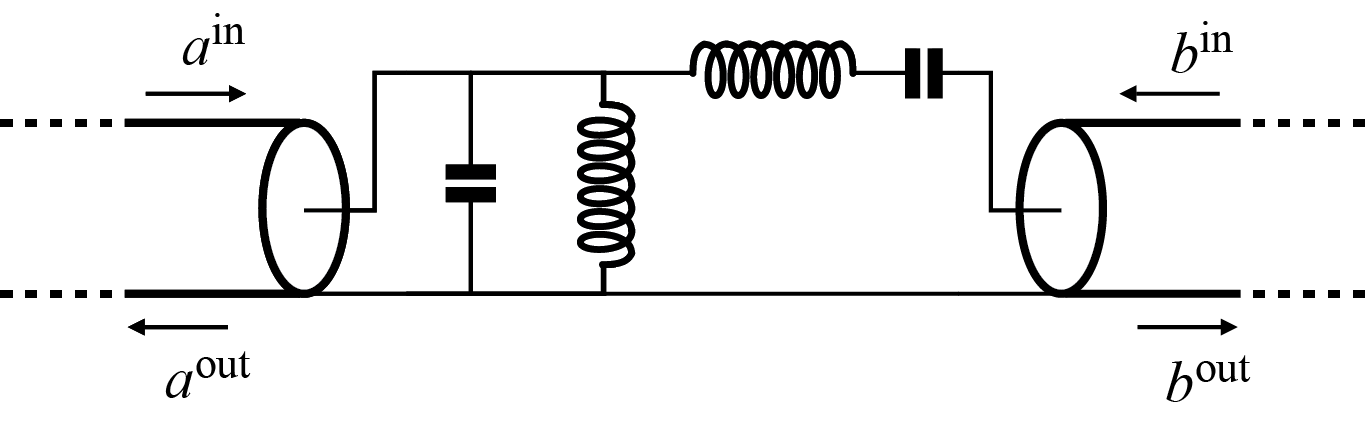}
\caption{\label{Fig_4} Example of a two-mode,
two-port circuit in which care must be taken in the amplitude factors of the
Quantum Langevin Equation.}
\end{figure}

Apart from a simple extension of the number of variables, this new equation
contains the rectangular matrix $\varepsilon _{MP}$ whose complex
coefficients are such that $\left\vert \varepsilon _{MP}\right\vert =1$.
This matrix can be computed from the details of the coupling of the lines to
particular elements of the circuit (capacitances or inductances, series or
parallel connections). A simple example of a situation where the $%
\varepsilon _{MP}$ cannot be set to unity by a re-definition of the mode
amplitude $\mathbf{a}_{M}$ is presented in Fig. \ref{Fig_4}. The general Input-Output Equation
takes the form%
\begin{equation}
\mathbf{a}_{M}=\sum_{P}\frac{1}{\sqrt{\kappa _{MP}}}\left[ \varepsilon
_{MP}\mathbf{a}_{M}^{\rm{in}}\left( t\right) +\varepsilon _{MP}^{-1}\mathbf{a}_{P}^{\rm{out}}\left(
t\right) \right] .
\end{equation}%


\section{Model amplifiers}
\label{sec_4}

\begin{figure}
\centering
\includegraphics[width = 0.9\textwidth]{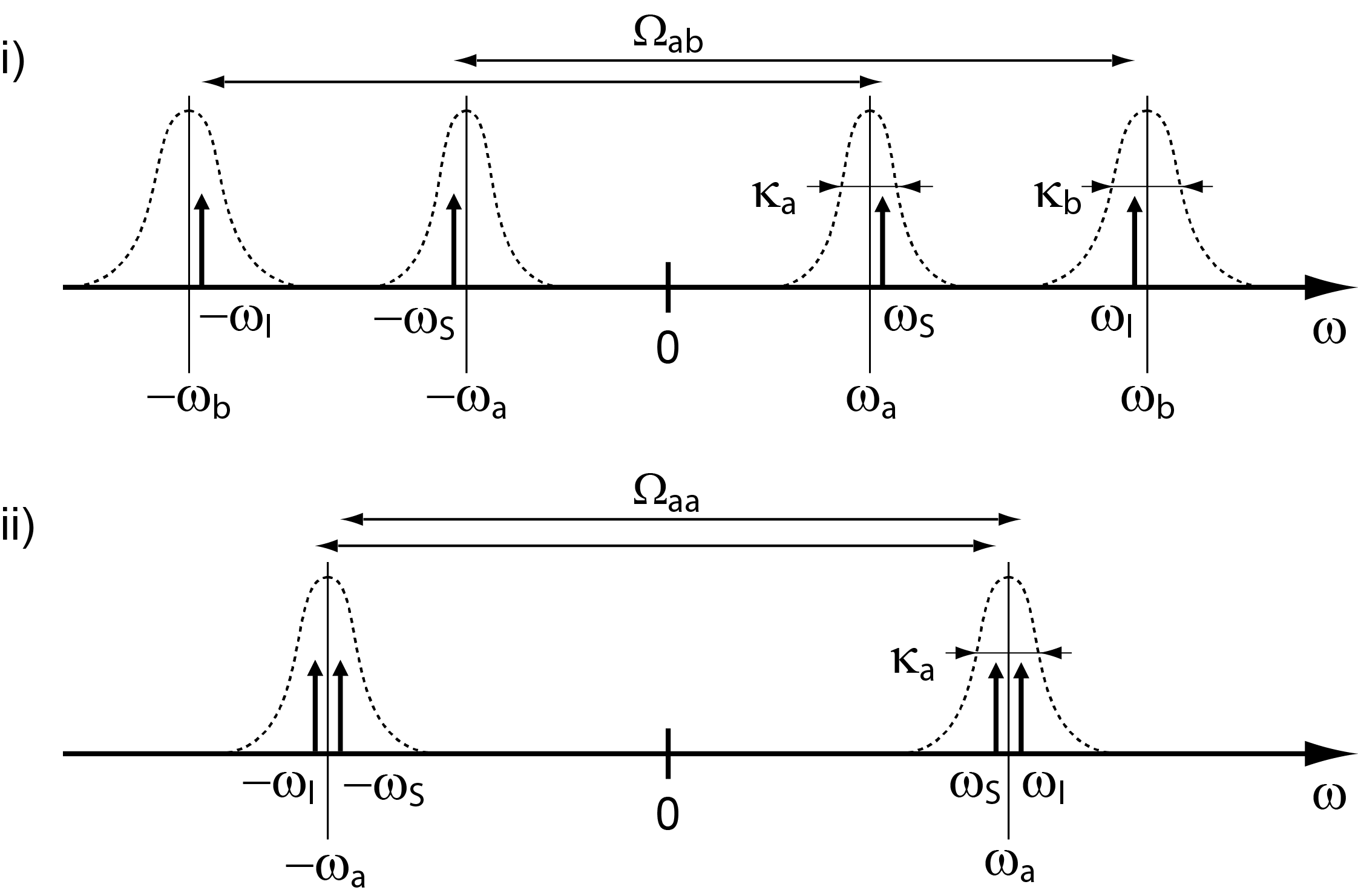}
\caption{\label{Fig_5} Frequency landscape for the non-degenerate (i) and degenerate (ii)
parametric amplifiers. The dashed lines corresponds to response curves of
each mode, as measured by a probe tone injected in the circuit elements of the mode. The vertical arrows
correspond to the spectral densities of the signal and the idler tones arriving in the circuit through its ports. The
horizontal arrows denote the frequency translations between signal and idler
operated by the parametrical modulation induced by the pump tone.}
\end{figure}

We have in the previous section introduced the Quantum Langevin Equation and
the Input-Output Equation which together allow, at least formally, to compute how a
given circuit processes signals propagating on transmission lines. We have
treated the simplest case of a 1-mode, 1-port circuit implementing an harmonic
oscillator circuit. However, the understanding of amplifiers starts by
considering necessarily a generalization of the simple harmonic case, namely
that of an effective, time-dependent quadratic Hamiltonian for the circuit.
We defer to the next section the discussion of how these Hamiltonians emerge
in practice from damped-driven non-linear systems. We will address first the
question of how an amplification function can arise from such emergent
effective quadratic form. Thus, let us consider the time-dependent effective
quadratic Hamiltonian%
\begin{equation}
\frac{\mathbf{H}}{\hbar }=\sum_{m}\omega _{m}\mathbf{a}_{m}^{\dag }\mathbf{a}_{m}+i\sum_{m\leq
p}g_{mp}^{\rm{eff}}\left( \mathbf{a}_{m}\mathbf{a}_{p}e^{i\left( \Omega _{mp}t+\theta _{mp}\right)
}-h.c.\right) +\text{neglected\ terms},
\end{equation}%
where $m$ and $p$ are circuit mode indices, and where the real, positive
parameters $\omega _{m}$ and $g_{ml}^{\rm{eff}}$ are in general functions
of elementary parameters of the circuit combined with the values of
time-dependent driving fields imposed from the outside and treated
classically. These driving fields excite the circuit, thus providing energy
for the amplification process and are often nicknamed ``pumps". We suppose
that there are ports that couple the modes $m$ to the outside world. In the
case of one port per mode, this coupling is described by constants $\kappa
_{m}$. The phase factors $e^{i\theta _{mp}}$ depend on the details of the
excitation, while the drive frequencies $\Omega _{mp}$ are in the vicinity
of $\omega _{m}+\omega _{p}$ (or sometimes $\left\vert \omega
_{m}-\omega _{p}\right\vert $, but we will not deal with this
case here). By vicinity, we mean within the bandwidth determined by the port
coupling constants: $\left\vert \Omega _{mp}-\omega _{m}-\omega
_{p}\right\vert  \leq \kappa _{m}\kappa _{p}/\left( \kappa
_{m}+\kappa _{p}\right)$. Here, we will limit
ourselves to two elementary cases: i) the simple two-port, 2-mode,
non-degenerate parametric amplifier with Hamiltonian%
\begin{equation}
\frac{\mathbf{H}^{\rm{NDPA}}}{\hbar }=\omega _{a}\mathbf{a}^{\dag }\mathbf{a}+\omega _{b}\mathbf{b}^{\dag
}\mathbf{b}+ig_{ab}\left( \mathbf{ab}e^{i\left( \Omega _{ab}t+\theta \right)
}-h.c.\right) 
\end{equation}%
and port coupling constants $\kappa _{a}$ and $\kappa _{b}$, and ii) the
simple one-port, 1-mode, degenerate parametric amplifier with Hamiltonian%
\begin{equation}
\frac{\mathbf{H}^{\rm{DPA}}}{\hbar }=\omega _{a}\mathbf{a}^{\dag }\mathbf{a}+ig_{aa}\left(
\mathbf{a}^{2}e^{i(\Omega _{aa}t+\theta)}-h.c.\right). 
\end{equation}%
In this case, there is a single port coupling constant $\kappa _{a}$. The
frequency landscapes corresponding to the two cases are represented
schematically on Fig. \ref{Fig_5}. 

From the Langevin Equation [Eq. \eqref{Quantum_Langevin}] and the Input-Output Equation [Eq. \eqref{input-output}] of the last section, we obtain for
non-degenerate case (i) the pair of equations:
\begin{eqnarray}
F(\omega_a, \kappa_a)\mathbf{a}^{\rm{out}} +g_{ab}e^{-i(\Omega _{ab}t+\theta)}{\mathbf{b}^{\rm{out}}}^{\dag } &=&-F(\omega_a, -\kappa_a)\mathbf{a}^{\rm{in}}-g_{ab}e^{-i(\Omega _{ab}t + \theta)}{\mathbf{b}^{\rm{in}}}^{\dag },\nonumber \\
F(\omega _{b}, \kappa _{b}) \mathbf{b}^{\rm{out}} +g_{ab}e^{-i(\Omega _{ab}t+\theta)}{\mathbf{a}^{\rm{out}}}^{\dag }&=&-F(\omega _{b}, -\kappa _{b}) \mathbf{b}^{\rm{in}} -g_{ab}e^{-i(\Omega _{ab}t+\theta)}{\mathbf{a}^{\rm{in}}}^{\dag }, \nonumber
\end{eqnarray}
where $F(\omega, \kappa) = d/dt + i\omega + \kappa/2$. For degenerate case (ii) there is only one equation:
\begin{eqnarray}
F(\omega _{a}, \kappa _{a}) \mathbf{a}^{\rm{out}} +2
g_{aa}e^{-i(\Omega _{aa}t+\theta)}{\mathbf{a}^{\rm{out}}}^{\dag }&=&-F(\omega _{a}, -\kappa _{a})\mathbf{a}^{\rm{in}} -2g_{aa}e^{-i(\Omega _{aa}t+\theta)}{\mathbf{a}^{\rm{in}}}^{\dag }\nonumber.
\end{eqnarray}%
Going to the Fourier domain and solving for outgoing waves as a function of
the incoming waves we find for case (i)%
\begin{equation}
\left[ 
\begin{array}{c}
\mathbf{a}^{\rm{out}}\left[ +\omega _{S}\right]  \\ 
\mathbf{a}^{\rm{out}}\left[ -\omega _{S}\right]  \\ 
\mathbf{b}^{\rm{out}}\left[ +\omega _{I}\right]  \\ 
\mathbf{b}^{\rm{out}}\left[ -\omega _{I}\right] 
\end{array}%
\right] =\left[ 
\begin{array}{cccc}
r_{SS} & 0 & 0 & s_{SI} \\ 
0 & r_{SS}^{\ast } & s_{SI}^{\ast } & 0 \\ 
0 & s_{IS}^{\ast } & r_{II}^{\ast } & 0 \\ 
s_{IS} & 0 & 0 & r_{II}%
\end{array}%
\right] \left[ 
\begin{array}{c}
\mathbf{a}^{\rm{in}}\left[ +\omega _{S}\right]  \\ 
\mathbf{a}^{\rm{in}}\left[ -\omega _{S}\right]  \\ 
\mathbf{b}^{\rm{in}}\left[ +\omega _{I}\right]  \\ 
\mathbf{b}^{\rm{in}}\left[ -\omega _{I}\right] 
\end{array}%
\right], 
\end{equation}%
where $\omega _{S}, \omega _{I}$ are the two signal and image (or
idler) frequencies, respectively, linked precisely by $\omega _{S}+\omega
_{I}=\Omega _{ab}$. Unlike in the case of simple harmonic circuits, an
input signal at one frequency can here be processed into an output signal at
another frequency. There is also a change in sign of the frequency in this
process, which is called phase conjugation, and this is why we represent the
scattering by a $4\times 4$ matrix. That matrix can be separated into two
blocks related by the simple complex conjugation relating Fourier
coefficients with opposite frequencies, a mathematical operation independent
of the physical phenomenon of phase conjugation. Phase conjugation manifests
itself practically in the following manner: if one advances the phase of the
input signal by a given quantity, the phase of the conjugated output signal
becomes retarded by the same quantity. The elements of the scattering matrix
are given by 
\begin{eqnarray}
r_{SS} &=&\frac{\chi _{a}^{-1}\left( \omega _{S}\right) ^{\ast }\chi
_{b}^{-1}\left( \omega _{I}\right) ^{\ast }+\ \rho _{ab}^{2}}{\chi
_{a}^{-1}\left( \omega _{S}\right) \chi _{b}^{-1}\left( \omega _{I}\right)
^{\ast }-\rho _{ab}^{2}}, \\
r_{II} &=&\frac{\chi _{a}^{-1}\left( \omega _{S}\right) \chi _{b}^{-1}\left(
\omega _{I}\right) +\rho _{ab}^{2}}{\chi _{a}^{-1}\left( \omega _{S}\right)
\chi _{b}^{-1}\left( \omega _{I}\right) ^{\ast }-\rho _{ab}^{2}}, \\
s_{SI} &=&\frac{2\rho _{ab}e^{-i\theta }}{\chi _{a}^{-1}\left( \omega
_{S}\right) \chi _{b}^{-1}\left( \omega _{I}\right) ^{\ast }-\rho _{ab}^{2}},
\\
s_{IS} &=&\frac{2\rho _{ab}e^{i\theta }}{\chi _{a}^{-1}\left( \omega
_{S}\right) \chi _{b}^{-1}\left( \omega _{I}\right) ^{\ast }-\rho _{ab}^{2}}.
\end{eqnarray}%
These expressions contain two ingredients: the single mode bare
susceptibilites $\chi $, which are given by%
\begin{equation}
\chi _{m}\left( \omega \right) =\frac{1}{1-2i\left( \omega -\omega
_{m}\right) /\kappa _{m}}
\end{equation}%
and the reduced effective mode coupling given by%
\begin{equation}
\rho _{ab}=\frac{2g_{ab}}{\sqrt{\kappa _{a}\kappa _{b}}}.
\end{equation}%
Its modulus squared is often called the mode cooperativity.
When the drive tone of the amplifier is optimally tuned $\Omega _{ab}=\omega _{a}+\omega _{b}$ 
and when monochromatic input signals are on resonance with their
corresponding mode $\omega _{S} =\omega _{a}, \omega _{I} =\omega _{b}$
the scattering matrix takes the simpler form%
\begin{equation}
\left[ 
\begin{array}{cccc}
\sqrt{G_{0}} & 0 & 0 & \sqrt{G_{0}-1}e^{-i\theta } \\ 
0 & \sqrt{G_{0}} & \sqrt{G_{0}-1}e^{i\theta } & 0 \\ 
0 & \sqrt{G_{0}-1}e^{-i\theta } & \sqrt{G_{0}} & 0 \\ 
\sqrt{G_{0}-1}e^{i\theta } & 0 & 0 & \sqrt{G_{0}}%
\end{array}%
\right],
\end{equation}%
where the zero-detuning, optimal amplifier power gain $G_{0}$ is%
\begin{equation}
G_{0}=\left( \frac{1+\rho _{ab}^{2}}{1-\rho _{ab}^{2}}\right) ^{2}.
\end{equation}%
It can be shown that the stability of the amplifier requires that $\rho
_{ab}<1$, i.e. there is a ceiling to the effective coupling between modes of
the circuit, beyond which amplification turns into spontaneous parametric oscillations.

Note that the determinant of the scattering matrix is unity even in the
fully general case. Also, it is important to realize that, quite generally,
the scattering is not reciprocal. A wave going from port \textit{b} to port 
\textit{a} acquires a phase factor $e^{-i\theta }$ from the drive which is
conjugate to the phase factor $e^{i\theta }$ accompanying the scattering
from port a to port b.

We now turn to case ii) in which the scattering relations, while involve
only one port, still are expressed as a $4\times 4$ matrix:%
\begin{equation}
\left[ 
\begin{array}{c}
\mathbf{a}^{\rm{out}}\left[ +\omega _{S}\right]  \\ 
\mathbf{a}^{\rm{out}}\left[ -\omega _{S}\right]  \\ 
\mathbf{a}^{\rm{out}}\left[ +\omega _{I}\right]  \\ 
\mathbf{a}^{\rm{out}}\left[ -\omega _{I}\right] 
\end{array}%
\right] =\left[ 
\begin{array}{cccc}
r_{SS} & 0 & 0 & s_{SI} \\ 
0 & r_{SS}^{\ast } & s_{SI}^{\ast } & 0 \\ 
0 & s_{IS}^{\ast } & r_{II}^{\ast } & 0 \\ 
s_{IS} & 0 & 0 & r_{II}%
\end{array}%
\right] \left[ 
\begin{array}{c}
\mathbf{a}^{\rm{in}}\left[ +\omega _{S}\right]  \\ 
\mathbf{a}^{\rm{in}}\left[ -\omega _{S}\right]  \\ 
\mathbf{a}^{\rm{in}}\left[ +\omega _{I}\right]  \\ 
\mathbf{a}^{\rm{in}}\left[ -\omega _{I}\right] 
\end{array}%
\right] .
\end{equation}%
Now the different frequencies are carried on the same port and are all in
the vicinity of the single resonance of the unique mode. Nevertheless, the
scattering coefficients are in the analytic continuation of the previous
expressions, in which one simply sets $\chi _{b}\left( \omega \right) =\chi
_{a}\left( \omega \right) $. A simplification occurs if the drive frequency
is precisely tuned to twice the effective resonant frequency, i.e. $\Omega
_{aa}=2\omega _{a}$, in which case $\chi _{a}\left( \omega _{I}\right)
=$ $\chi _{a}\left( \omega _{S}\right) ^{\ast }$ and the subblock of the
scattering matrix takes the form%
\begin{equation}
\left[ 
\begin{array}{c}
\mathbf{a}^{\rm{out}}\left[ +\omega _{S}\right]  \\ 
\mathbf{a}^{\rm{out}}\left[ -\omega _{I}\right] 
\end{array}%
\right] =\frac{1}{D}\left[ 
\begin{array}{cc}
\left\vert \chi _{a}^{-1}\left( \omega _{S}\right) \right\vert ^{2}+\ \rho
_{aa}^{2} & 2\rho _{aa}e^{-i\theta } \\ 
2\rho _{aa}e^{i\theta } & \left\vert \chi _{a}^{-1}\left( \omega
_{S}\right) \right\vert ^{2}+\ \rho _{aa}^{2}%
\end{array}%
\right] \left[ 
\begin{array}{c}
\mathbf{a}^{\rm{in}}\left[ +\omega _{S}\right]  \\ 
\mathbf{a}^{\rm{in}}\left[ -\omega _{I}\right] 
\end{array}%
\right],
\end{equation}%
with $D=\left[ \chi _{a}^{-1}\left( \omega _{S}\right) \right] ^{2}-\ \rho
_{aa}^{2}$ where $\rho _{aa}=\frac{4g_{aa}}{\kappa _{a}}$. We also introduce the in-phase and quadrature components of the
incoming and outgoing waves:%
\begin{eqnarray}
\mathbf{a}_{\Vert, \bot }^{\rm{in, out}}\left( \delta \omega \right)  &=&\mathbf{a}^{\rm{in, out}}\left[ \omega
_{S}\right] \pm e^{-i\theta }\mathbf{a}^{\rm{in, out}}\left[ -\omega _{I}\right], 
\end{eqnarray}%
where $\delta \omega =\omega _{S}-\omega _{a}=\omega _{a}-\omega
_{I}$. The meaning of this transformation can be illustrated by the
following consideration, which supposes $\theta =0$ for simplicity.
Classically, if a signal is such that 
\begin{equation}
y\left( t\right) =f\left( t\right) \cos \left( \omega _{a}t\right)
+g\left( t\right) \sin \left( \omega _{a}t\right),
\end{equation}%
with in-phase and quadrature modulation components $f\left( t\right) $ and $%
g\left( t\right) $ slow compared to $\left( \omega _{a}\right) ^{-1}$,
then 
\begin{eqnarray}
y_{\Vert }\left( \delta \omega \right)  =f\left( \delta \omega \right),\hspace{1cm}y_{\bot }\left( \delta \omega \right) =g\left( \delta \omega \right).
\end{eqnarray}%
One can easily check that the effect of the angle $\theta $ associated with
the time dependence of the effective Hamiltonian is just to rotate the
component signals in the Fresnel plane. In the representation where the
in-phase and quadrature components form the basis signals, we find that the
scattering matrix is diagonal 
\begin{eqnarray}
\mathbf{a}_{\Vert }^{\rm{out}}\left( \delta \omega \right)  &=&\frac{\left\vert \chi
_{a}^{-1}\left( \omega _{S}\right) \right\vert ^{2}+2\rho _{aa}+\ \rho
_{aa}^{2}}{D}\mathbf{a}_{\Vert }^{\rm{in}}\left( \delta \omega \right) =\Lambda _{\Vert
}\left( \delta \omega \right) \mathbf{a}_{\Vert }^{\rm{in}}, \\
\mathbf{a}_{\bot }^{\rm{out}}\left( \delta \omega \right)  &=&\frac{\left\vert \chi
_{a}^{-1}\left( \omega _{S}\right) \right\vert ^{2}-2\rho _{aa}+\ \rho
_{aa}^{2}}{D}\mathbf{a}_{\bot }^{\rm{in}}\left( \delta \omega \right) =\Lambda _{\bot
}\left( \delta \omega \right) \mathbf{a}_{\bot }^{\rm{in}}.
\end{eqnarray}%
The property of the scattering matrix to have unity determinant imposes%
\begin{equation}
G_{\Vert }G_{\bot }=1.
\end{equation}%
where $G_{\Vert ,\bot }=\left\vert \Lambda _{\Vert ,\bot }\left( \delta
\omega \right) \right\vert ^{2}$. Thus, in this mode of operation of the
degenerate parametric amplifier, one quadrature of the signal is amplified
while the other is de-amplified. If the input signal consists only of vacuum
fluctuations, the amplifier squeezes these fluctuations for one quadrature,
making it less uncertain than the so-called standard quantum limit, which is
associated to a standard deviation corresponding to the square root of a
quarter of a photon (the half photon of the zero-point motion is split
evenly between the two quadratures, and only one is squeezed) \cite{Drummond_Ficek_2013}.

In the non-degenerate case (i) a more complex form of squeezing -- two-mode
squeezing -- occurs in the four dimensional phase space of the quadratures
of the two propagating signals incident on the circuit \cite{Drummond_Ficek_2013}.

The non-degenerate parametric amplifier is usually employed as a sort of RF
op-amp: the idler port is connected to a cold matched load emulating an
infinite transmission line at zero-temperature and the device viewed from
the signal port functions as a reflection amplifier operating in the phase
preserving mode: for signals having a bandwidth small compared to that of
the amplifier
\begin{equation}
\mathbf{a}^{\rm{out}}=\sqrt{G}\left(\mathbf{a}^{\rm{in}}+\sqrt{1-\frac{1}{G}}\mathbf{b}^{\rm{in}\dag }\right).
\label{simple_gain_input-output_expression}
\end{equation}%
The second term on the right of this last expression shows that quantum
noise entering through the \textit{b} port must necessarily be added to the
amplified signal \cite{Caves_1982}. This added noise contribution amounts, in the large gain
limit $G\gg 1$ and for an idler port at zero temperature, to a half-photon
at the signal frequency, referred to the input. It can be seen as an evil necessary to preserve the commutation relation%
\begin{equation}
\left[ \mathbf{a}^{\rm{out}},\mathbf{a}^{\rm{out}\dag }\right] =\left[ \mathbf{a}^{\rm{in}},\mathbf{a}^{\rm{in}\dag }\right].
\end{equation}%
More practically, the extra half-photon of noise can also be seen as a
consequence of the Heisenberg Uncertainty Principle. A phase preserving
amplifier processes equally both quadratures, which in quantum mechanics are
non-commuting observables. Since the process of amplification is equivalent
to measurement, the extra noise forbids that both quadratures are known
precisely simultaneously, in accordance with the central principle of
quantum mechanics. An amplifier functioning in this Heisenberg regime where
the efficiency of the amplification process is only limited by irreducible
quantum fluctuations is said to be quantum-limited.

\section{Practical amplifier circuits based on Josephson junction circuits}
\label{sec_5}

We have introduced in the previous section the notion of effective,
time-dependent quadratic Hamiltonians and shown how their generic form lead
to the amplification of a quantum signal with a noise limited solely by the
Heisenberg Uncertainty Principle. We now explain in this section how such
quadratic forms can arise from a practical non-linear, damped and driven
system based on Josephson tunnel junctions.

\begin{figure}
\centering
\includegraphics[width=\textwidth]{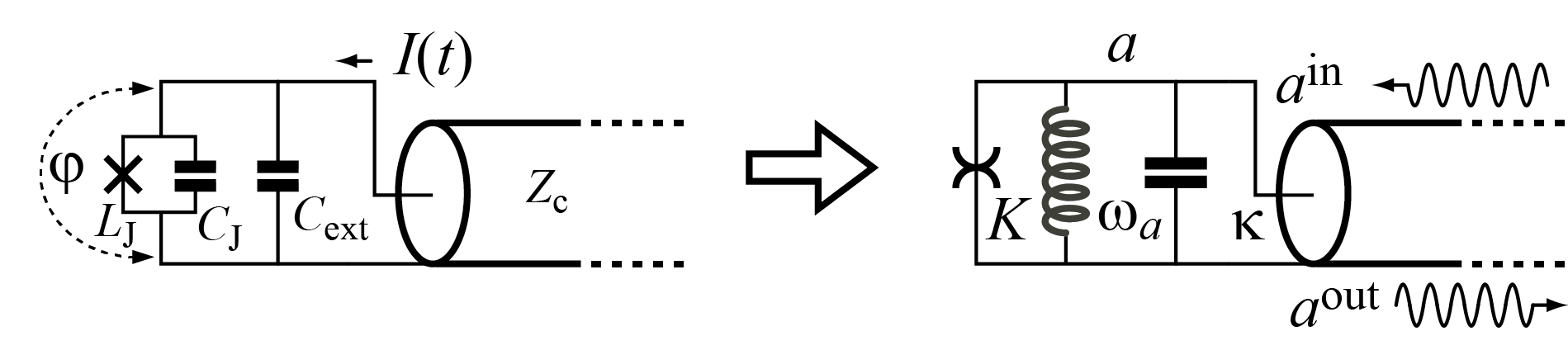}
\caption{\label{Fig_6}1-mode, 1-port Josephson amplifier involving only the Josephson
inductance. Left panel is schematic of Josephson tunnel junction, itself
consisting of a Josephson tunnel element playing the role of non-linear
inductance (cross, $L_{J}$) and a junction capacitance $C_{J}$, in parallel
with an external capacitance $C_{\rm{ext}}$ and a transmission line bringing in
the current $I(t)$. The variable $\protect\varphi $ is the phase across the
junction. On the right panel, simplified schematic based on an RWA treatment
where the oscillator is reduced to its frequency $\protect\omega _{a}$ and
damping rate $\protect\kappa $, with the Josephson non-linearity manifesting
itself as a simple Kerr component (opposing arcs symbols) characterized by the parameter K, the shift in frequency of the oscillator corresponding to 1 photon. The degree of
freedom is described by the standing photon ladder operator $\mathbf{a}.$}
\end{figure}


\subsection{Driven microwave oscillator whose inductance is a single
Josephson element: Duffing-like dynamics}

We first examine the simplest case of a 1-mode, 1-port circuit in which the
inductance is just the Josephson element of superconducting tunnel junction
(see Fig. \ref{Fig_6}) \cite{Siddiqi_Devoret_2004, Vijay_2008}. The Hamiltonian of such systems is given by%
\begin{eqnarray}
\mathbf{H} &=&\mathbf{H}_{\rm{circ}}-\frac{\hbar }{2e}\bm{\varphi}\cdot \mathbf{I}+\text{ 
}\mathbf{H}_{\rm{env}}, \\
\mathbf{H}_{\rm{circ}} &=&-E_{J}\cos \bm{\varphi}+\frac{\mathbf{Q}^{2}}{2C_{\Sigma }},
\end{eqnarray}%
where $E_{J}=\left( \frac{\hbar }{2e}\right) ^{2}/L_{J}$ is the Josephson
energy, $C_{\Sigma }=C_J+C_{\rm{ext}}$ is the total capacitance in parallel
with the Josephson element,  $\bm{\varphi}$ the gauge-invariant phase
difference across the junction, $\mathbf{Q}$ the charge conjugate to the phase $%
\left[ \bm{\varphi},\mathbf{Q}\right] =2ei$, $\mathbf{H}_{\rm{env}}$ the Hamiltonian
of the transmission line, including the pump arriving through this channel
and $\mathbf{I}$ the current operator belonging to the degrees of freedom of
the line. The amplifier functions with $\langle\bm{\varphi}\rangle$ having excursions
much less than $\pi /2$ and the cosine function in the Hamiltonian can be
expanded to $4^{th}$ order only, with the $\bm{\varphi}^{4}$ term treated as a
perturbation \cite{Manucharyan_Devoret_2007, Nigg_Girvin_2012}. Introducing the ladder operators of the
single mode of the circuit%
\begin{equation}
\bm{\varphi}=\varphi _{\rm{ZPF}}\left( \mathbf{a}+\mathbf{a}^{\dag }\right) 
\end{equation}%
and working in the framework of both an expansion in $\varphi _{\rm{ZPF}}=\left(
2e^2/\hbar\right) ^{1/2}\left( L_{J}/C_{\Sigma }\right) ^{1/4}$ and RWA, the
hamiltonian of the circuit simplifies to%
\begin{equation}
\frac{\mathbf{H}_{\rm{circ}}}{\hbar }=\tilde{\omega} _{a}\mathbf{a}^{\dag }\mathbf{a}+\frac{K}{2}\mathbf{a}^{\dag
}\mathbf{a}\left( \mathbf{a}^{\dag }\mathbf{a}-1\right), 
\end{equation}%
where, in the regime $\varphi _{\rm{ZPF}}\ll 1$, $K=-e^{2}/(2\hbar C_{\Sigma })$ and $%
\tilde{\omega} _{a}=1/\sqrt{L_{J}C_{\Sigma }} + K$. The Quantum Langevin Equation
applied to this system yields%
\begin{equation}\label{qle_jba}
\frac{d}{dt}\mathbf{a}=-i\left( \tilde{\omega} _{a}+K\mathbf{a}^{\dag }\mathbf{a}\right) \mathbf{a}-\frac{\kappa }{2}\mathbf{a}+%
\sqrt{\kappa }\mathbf{a}^{\rm{in}}\left( t\right),
\end{equation}%
where $\tilde{\omega}_{a}\gg \kappa _{a}\gg K$. This last equation is the quantum
version, in the RWA approximation, of the equation describing systems
modeled by the Duffing equation. The classical Duffing oscillator obeys the
equation 
\begin{equation}
m\ddot{x}+\eta \dot{x}+m\omega _{0}^{2}x(1+\mu x^{2})=f_{D}\cos \left(
\omega _{D}t\right) +f_{P}\left( t\right) 
\end{equation}%
for the position variable $x$ having mass $m$, small amplitude spring
constant $m\omega _{0}^{2}$, friction coefficient $\eta $ and driven at
frequency $\omega _{D}$, which is close to the small amplitude resonant
frequency $\omega _{0}$. A small probe force $f_{P}\left( t\right) $ allows
to study the displacement response of the system. Non-linearity of the
oscillator corresponds here to the spring constant being dependent
quadratically on position.

Here, for our amplifier, $\kappa $ plays the role of the damping rate $\eta
/m$, and $K$ plays the role of $\mu $. Let us now suppose that, in addition
to the signal to be processed, the $a$ port also receives an intense drive
tone described by a propagating coherent state with amplitude $\alpha ^{\rm{in}}$
and frequency $\Omega $. We treat this drive by the change of variable%
\begin{eqnarray}
\mathbf{a}^{\rm{in}}\left( t\right)  &=&\alpha ^{\rm{in}}e^{-i\Omega t}+ \mathbf{\bm\delta a}^{\rm{in}}\left(
t\right), \\
\mathbf{a}\left( t\right)  &=&\alpha e^{-i\Omega t}+\mathbf{\bm\delta a}\left( t\right) .
\end{eqnarray}%
We aim to solve for the semi-classical amplitude $\alpha$ from Eq. \eqref{qle_jba}:
\begin{equation}
\frac{d\alpha}{dt} - i\Omega\alpha=-i\tilde{\omega} _{a}\alpha-iK|\alpha|^2\alpha-\frac{\kappa _{a}}{2}\alpha+\sqrt{%
\kappa _{a}}\alpha ^{\rm{in}}.
\end{equation}%
By treating the non-linear term as a perturbation, we obtain the
self-consistent algebraic equation in steady state:%
\begin{equation}
\alpha =\frac{i\sqrt{\kappa _{a}}\alpha ^{\rm{in}}}{\left( \Omega -\tilde{\omega}
_{a}\right) +\frac{i\kappa _{a}}{2}-K\left\vert \alpha \right\vert ^{2}},
\end{equation}%
which in general yields for the c-number $\alpha $ a complex value such%
\begin{eqnarray}
\left\vert \alpha -\alpha _{0}\right\vert\ll1,  \hspace{0.5cm}
\alpha _{0} =\frac{i\sqrt{\kappa _{a}}\alpha ^{\rm{in}}}{\left( \Omega
-\tilde{\omega} _{a}\right) +\frac{i\kappa _{a}}{2}-4K\left\vert \alpha
^{\rm{in}}\right\vert ^{2}/\kappa _{a}}.
\end{eqnarray}%
This value leads to the effective Hamiltonian for the degenerate parametric
amplifier arising from the pumping of the Josephson junction%
\begin{equation}
\frac{\mathbf{H}}{\hbar }=\omega _{a}\bm{\delta} \mathbf{a}^{\dag }\bm{\delta}\mathbf{a}+\left[
g_{aa}e^{i\left( \Omega _{aa}t+\theta \right) }\left( \bm{\delta}\mathbf{a}\right)
^{2}+h.c.\right] 
\end{equation}%
with%
\begin{eqnarray}
\label{eq_a}
\omega _{a} &=&\tilde\omega _{a}+2K\left\vert \alpha \right\vert
^{2}, \\
g_{aa}e^{i\theta } &=&K\alpha ^{\ast 2}, \\\label{eq_b}
\Omega _{aa} &=&2\Omega.
\end{eqnarray}%
Eqs. \eqref{eq_a}, \eqref{eq_b} put into light two drawbacks of this type of amplifier: the center frequency of the band of the amplifier shifts as the pump amplitude is increased and the pump tone needs to be at the center of the band for optimal amplification.
The use of two pumps frequencies $\Omega _{1}$ and $\Omega _{2}$ such
that $\Omega _{aa}=\Omega _{1}+$ $\Omega _{2}$ facilitates the use of this
parametric amplifier \cite{Kamal_Devoret_2009}.  

The device has noticeable gain when $K\left\vert \alpha ^{\rm{in}}\right\vert
^{2}/\kappa _{a}^{2}$ is of order unity, implying that the number of pump
photons in the oscillator is of order $\kappa _{a}/K$, a large number by
hypothesis. This justifies our treatment of the pump drive as a c-number.
Neglected terms such as the non-RWA term $\left( \delta a\right)
^{3}e^{i\Omega t}$ have smaller factors and can themselves be treated as
perturbations on top of the standard degenerate parametric amplifier
formalism. It is worth noting that for this device, the pump tone and the
signal tone must enter the circuit on the same port, which is inconvenient given the widely different amplitude levels of these two waves. 

Amplifiers based on the same Duffing type of non-linearity can also be
fabricated with two-port circuits containing arrays of Josephson junctions \cite{Castellanos-Beltran_Lehnert_2008, Macklin_Siddiqi_2015}.

\subsection{A parametrically driven oscillator: the DC-SQUID driven by RF
flux variation}

\begin{figure}
\centering
\includegraphics[width=0.8\textwidth]{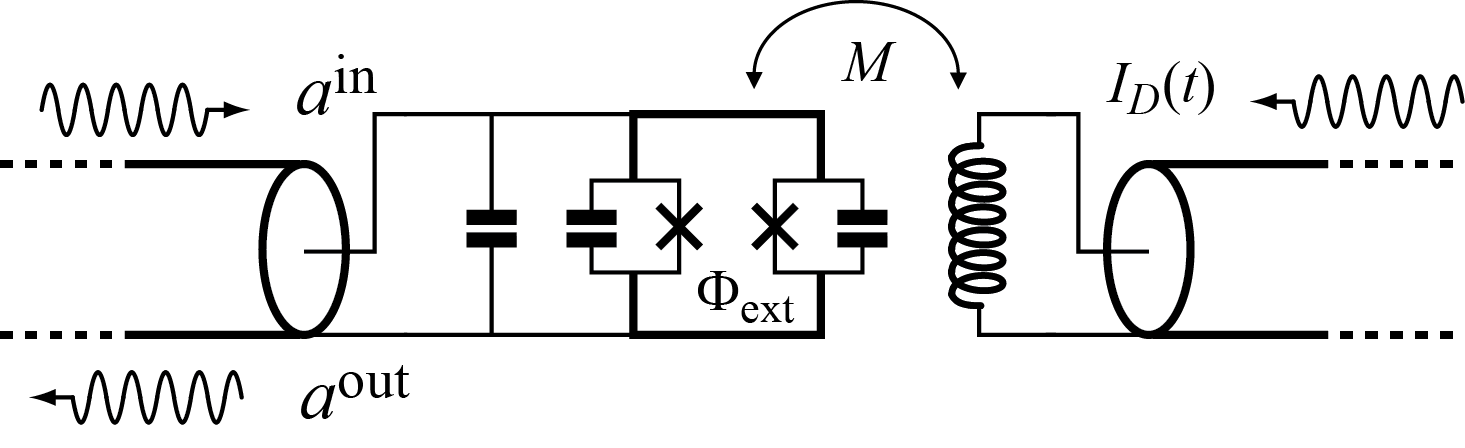}
\caption{\label{Fig_7}Parametrically driven oscillator
based on the property of the Josephson inductance of a DC-SQUID (two
Josephson junctions in parallel forming a loop, here represented by a
thicker line) to be modulated by the variation of an external flux $\Phi
_{ext}$. The modulation arises from an RF drive current $I_{D}\left(
t\right) =I_{D}^{RF}\cos \Omega t$ in the primary of a transformer that
creates though its mutual inductance $M$ a sinusoidal flux variation in the
loop of the DC-SQUID.}
\end{figure}

Another class of Josephson circuit implementing parametric amplifiers at
microwave frequencies is the RF-Flux-driven DC-SQUID (see Fig. \ref{Fig_7}).

It turns out that
this parametric drive can be implemented in Josephson circuits by taking a
DC-SQUID, which is formed by two nominally identical Josephson junctions in
parallel and modulating at the RF\ pump frequency the flux $\Phi _{ext}$
threading the superconducting loop between them (here the term DC refers to
the circulating current in the loop due to an external bias flux). One
exploits the functional form of the Josephson inductance of the DC-SQUID%
\begin{equation}
L_{J}^{SQUID}=\frac{L_{J}}{\cos \left\vert \pi \frac{\Phi _{ext}}{\Phi _{0}}%
\right\vert },
\end{equation}%
where $\Phi _{0}=h/2e$ is the flux quantum and $L_{J}/2$ is the Josephson
inductance of each individual junction. When%
\begin{equation}
\Phi _{ext}=\frac{\Phi _{0}}{4}\left[ 1+\varepsilon \cos \left( \Omega
t\right) \right],
\end{equation}%
with $\Omega $ close to the resonant frequency of the SQUID $1/\sqrt{%
C_{\Sigma }L_{J}^{SQUID}}$ and $\varepsilon \ll 1$, one implements the
parametrically driven harmonic oscillator with relative frequency modulation
parameter $\mu _{r}=\pi \varepsilon /4$. This modulation is produced by a
drive current $I_{D}\left( t\right) =I_{D}^{RF}\cos \left( \Omega t\right) $
at the primary of the transformer coupling the transmission line of an RF
pump to the flux of the SQUID (see Fig. \ref{Fig_7}).  Classically, the parametrically
driven harmonic oscillator obeys the equation%
\begin{equation}
m\ddot{x}+\eta \dot{x}+m\omega _{0}^{2}x\left[ 1+\mu _{r}\cos \left( \Omega
t\right) \right] =f_{P}\left( t\right).
\end{equation}%
In contrast with the Duffing oscillator above, this system is described by a
fully linear, albeit time-dependent, equation. The drive, instead of
appearing as a force coupled directly to position, now modulates the spring
constant with a relative amplitude $\mu _{r}$. The system behaves as an
amplifier when the argument of the cosine modulation term is such that the
drive frequency $\Omega $ is close to the resonant frequency $\omega _{0}$.
In the weak damping limit $\eta \ll m\omega _{0}$, the quantum version of
this oscillator is directly a one-port, one-mode system described by our
degenerate amplifier Hamiltonian%
\begin{equation}
\frac{\mathbf{H}}{\hbar }=\omega _{a}\mathbf{a}^{\dag }\mathbf{a}+\left[ g_{aa}e^{i\Omega
_{aa}t}\mathbf{a}^{2}+h.c.\right],
\end{equation}%
where%
\begin{eqnarray}
\Omega _{aa} &=&\Omega,  \\
g_{aa} &=&\mu _{r}\omega _{0}/4.
\end{eqnarray}

Note that now the drive frequency needs to be near twice the resonance
frequency of the amplified mode, unlike in the Duffing case, and it is thus
easier to decouple the pump tone from the weak signal to be amplified. This
type of amplifier has been implemented in several labs \cite{Yamamoto_Tsai_2008, Johansson_Nori_2009, Zhou_Esteve_2014, Mutus_Martinis_2014}.

\subsection{3-mode circuit employing the purely dispersive Josephson 3-wave
mixer}

We have just described the two ways in which a circuit involving one or two
Josephson junctions can implement the degenerate parametric amplifier (case
(ii) of last section). The non-degenerate parametric amplifier (case (i) of
last section) can be implemented by a 3-mode, 3-port circuit employing four
junctions forming the so-called Josephson ring modulator, a purely
dispersive 3-wave mixer (see details in \cite{Bergeal_Devoret_2010, Bergeal_Devoret_2010_a, Abdo_Devoret_2013_b}). 

\begin{figure}
\centering
\includegraphics[width=\textwidth]{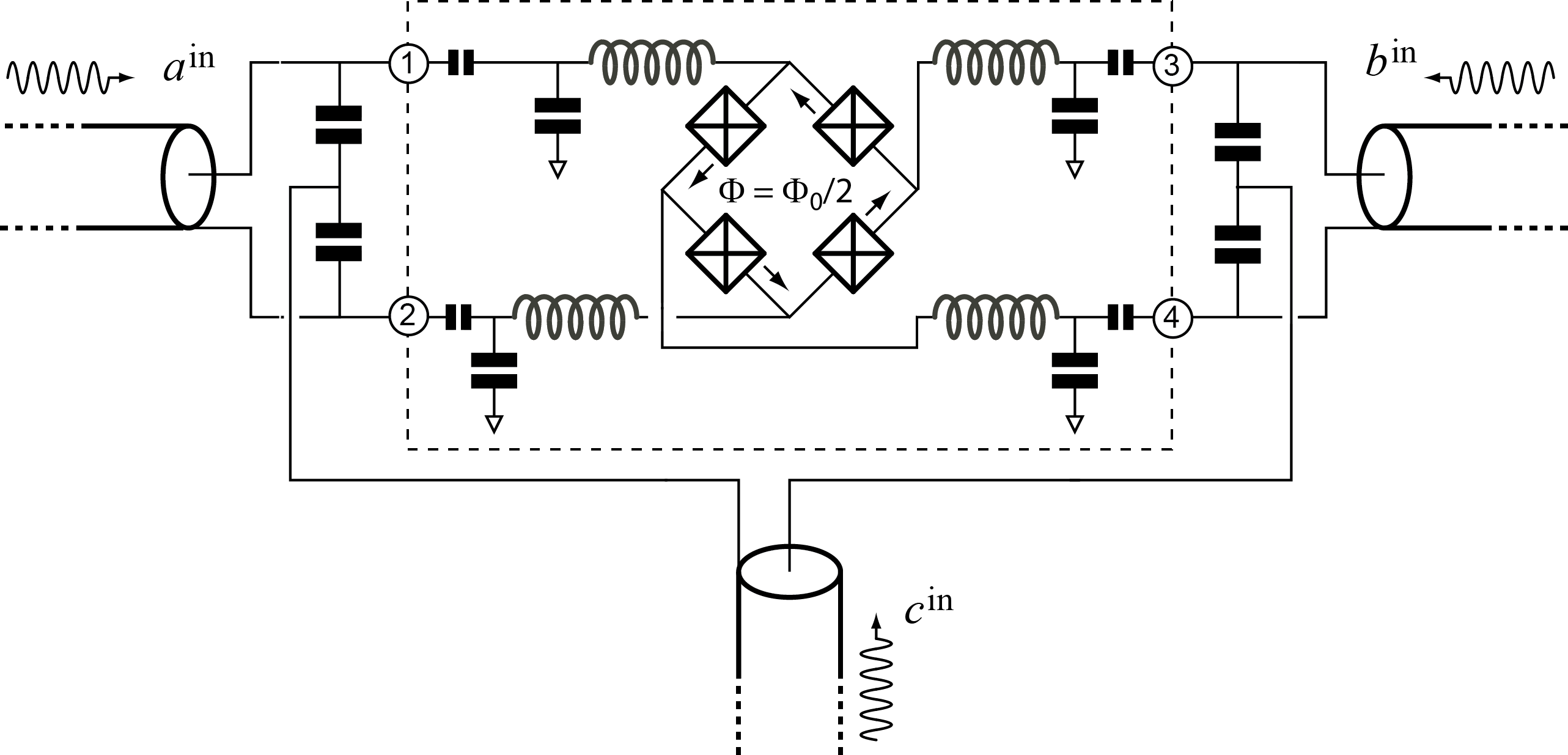}
\caption{\label{Fig_8}Schematic circuit of the
purely dispersive 3-wave mixer (dashed line) involving three microwave modes,
themselves coupled to three ports. The system functions as a non-degenerate
parametric amplifier with mode $a$ and $b$ playing the role of the signal 
and idler, while mode $c$ is used to couple in the pump tone. While there
are in principle four modes coupled by the junctions (cross inside a square,
denoting both the Josephson element and associated capacitance), the symmetry
of the circuit when the junctions are
identical, imposes that only three modes participate in the nonlinear interaction. A flux threading the ring of junctions induces a current (arrows) that replaces one of the four waves coupled by the junction.}
\end{figure}

Three microwave standing wave resonators are coupled by this
last element and are described by the Hamiltonian%
\begin{equation}
\frac{\mathbf{H}}{\hbar }=\omega _{a}\mathbf{a}^{\dag }\mathbf{a}+\omega _{b}\mathbf{b}^{\dag }\mathbf{b}+\omega
_{c}\mathbf{c}^{\dag }\mathbf{c}+g_{3}\left( \mathbf{a}+\mathbf{a}^{\dag }\right) \left( \mathbf{b}+\mathbf{b}^{\dag }\right)
\left( \mathbf{c}+\mathbf{c}^{\dag }\right),  \label{JPC_ideal_Hamiltonian}
\end{equation}%
together with their port coupling $\kappa _{a}$, $\kappa _{b}$ and $\kappa
_{c}$. The frequency scales are such that%
\begin{equation}
\omega _{c}\gg \omega _{b}>\omega _{a}>\kappa _{c}\gg \kappa _{a}\simeq
\kappa _{b}\gg g_{3}.
\end{equation}%
The trilinear coupling term, treated as a perturbation, possesses the
precious property that it does not, at the lowest order, offset the
frequency of the quadratic terms when the modes are occupied by coherent
signals, unlike the Kerr term above $\frac{K}{2}\mathbf{a}^{\dag }\mathbf{a}\left( \mathbf{a}^{\dag
}\mathbf{a}-1\right) $. Other terms of higher order have been neglected in the
Hamiltonian (\ref{JPC_ideal_Hamiltonian}). They ensure that the system
remains stable when the amplitudes become large, as the trilinear coupling
renders by itself the Hamiltonian unstable. In the regime where the $c$ mode
is driven by a large coherent field $\alpha _{c}^{\rm{in}}e^{-i\Omega t}$, we can
neglect the fluctuating part of the corresponding operator. The previous
Hamiltonian can be treated as%
\begin{equation}
\frac{\mathbf{H}^{\rm{eff}}}{\hbar }=\omega _{a}\mathbf{a}^{\dag }\mathbf{a}+\omega _{b}\mathbf{b}^{\dag
}\mathbf{b}+2g_{ab}\left( \mathbf{a}+\mathbf{a}^{\dag }\right) \left( \mathbf{b}+\mathbf{b}^{\dag }\right) \cos
\left( \Omega _{ab}t+\theta \right),
\end{equation}%
where%
\begin{eqnarray}
g_{ab}\cos \left( \Omega _{ab}t+\theta \right)  &=&g_{3}\Re\left[
\frac{\sqrt{\kappa _{c}}\alpha _{c}^{\rm{in}}e^{-i\Omega t}}{-i\left( \omega
_{ab}^{D}-\omega _{c}\right) +\kappa _{c}}\right],  \\
\Omega _{ab} &=&\Omega .
\end{eqnarray}%

\begin{figure}[!h]
\centering
\includegraphics[width=\textwidth]{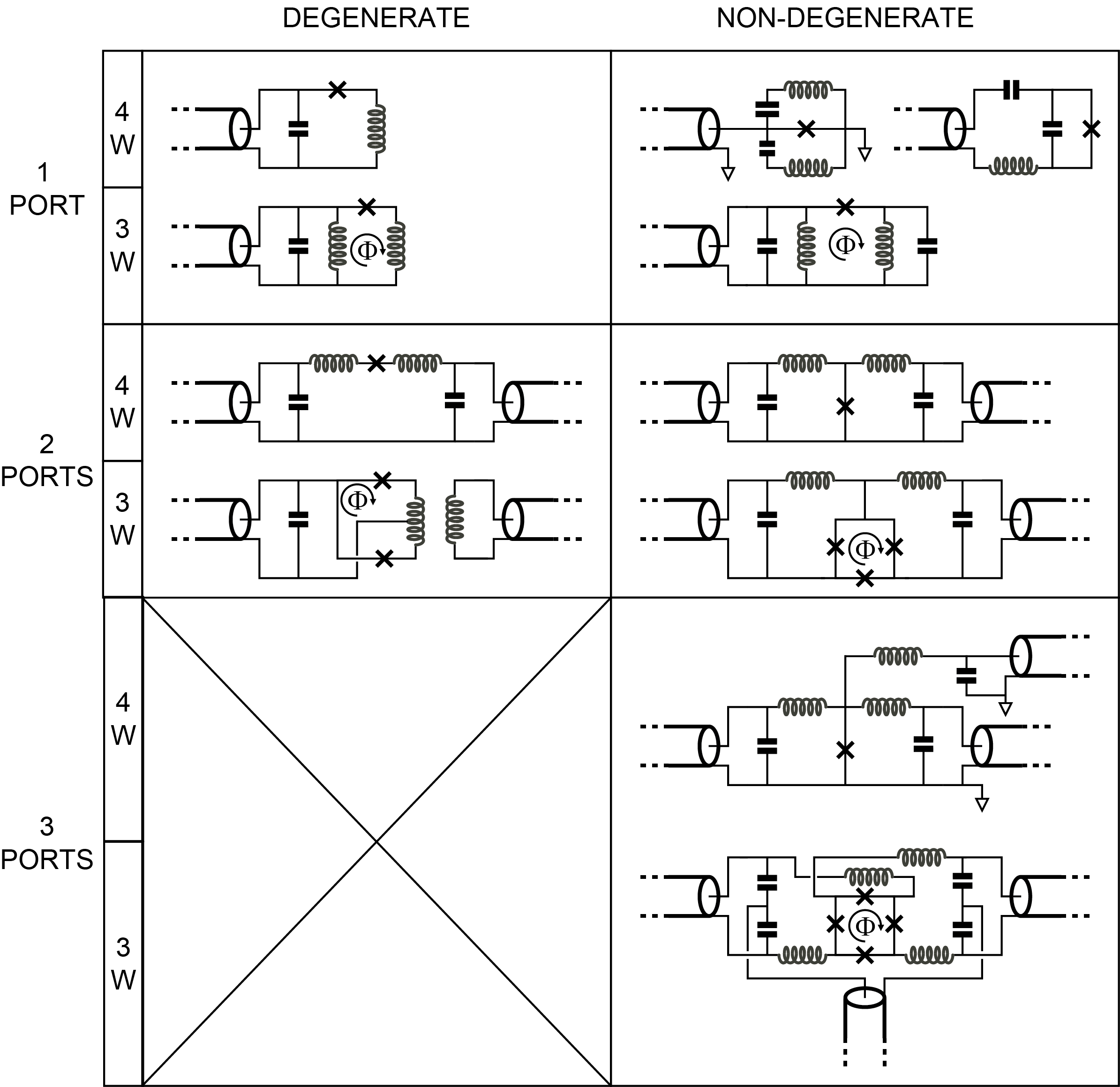}
\caption{\label{Fig_9} Minimalistic versions of the various Josephson circuits implementing parametric amplification of quantum signals. The circuits are classified according to the degenerate/non-degenerate character of the amplifying process (one or two standing modes). The four-wave (4W) or the three-wave (3W) labels characterize the mixing process  for the signal and idler waves taking place in the Josephson junctions. In the three-wave process, the place of one of the four-waves incident on the junction is replaced by a DC current generated by the externally applied flux $\Phi$. We also distinguish circuits by the number of ports through which the signal, idler and the pump waves are delivered. In the upper left corner (minimal complexity), the three waves are approximately at the same frequency and arrive through the same port, whereas in the lower right corner (maximal complexity), the three waves are both spatially and spectrally separated. In the upper right corner, we have represented two implementations of the 1-port, 4-wave, non-degenerate parametric amplifier. In the circuit on the left-handside, the two modes share a common junction but are symmetrically coupled to the port, whereas on the right-handside, the two modes are gauge-coupled and are asymmetrically coupled to the port \cite{Roy_Vijay_2015}.  }
\end{figure}

When one works within the framework of the Rotating Wave Approximation and $%
\Omega _{ab}\simeq \omega _{a}+\omega _{b}$, the fast rotating terms can be
neglected and one recovers the Hamiltonian of the generic non-degenerate
parametric amplifier%
\begin{equation}
\frac{\mathbf{H}^{\rm{NDPA}}}{\hbar }=\omega _{a}\mathbf{a}^{\dag }\mathbf{a}+\omega _{b}\mathbf{b}^{\dag }\mathbf{b}+\left[
g_{ab}\mathbf{ab}e^{i\left( \Omega _{ab}t+\theta \right) }+h.c.\right].
\end{equation}

\section{Concluding Summary and Perspectives}
\label{sec_6}
In this short review, we have outlined the different ways in which Josephson circuits can implement the functions of parametric amplification. The key organizing concept is the effective quadratic time-dependent Hamiltonian which comes in two forms: degenerate and non-degenerate, depending on whether the signal and idler waves occupy the same physical degree of freedom or two separate ones. In Fig. \ref{Fig_9}, we summarize the different circuit configurations leading, on one hand, to the degenerate case and, on the other hand, to the non-degenerate case (left and right columns, respectively). The figure also classifies circuits depending on the number of access ports, the simpler case being that of 1-port carrying the signal, idler and pump waves (upper left panels), while the case in which the signal, idler and the pump waves are separated in both temporally and spatially is shown in the bottom right panel. The circuit complexity increases when going from the upper left corner of Fig. \ref{Fig_9} to the lower right one. 

This survey of amplifiers has not covered the following topics of current interest: dynamic range \cite{Abdo_Devoret_2013_b}, directionality \cite{Kamal_Devoret_2011, Abdo_Devoret_2014, Macklin_Siddiqi_2015}, efficiency and frequency-conversion. Lastly, recently several ideas have been proposed to make the amplifier avoid the gain-bandwidth compromise and the instability at the onset of parametric oscillation \cite{Metelmann_Clerk_2014, Ranzani_Aumentado_2015}. These topics will be covered in an extended version of this review.

\section*{Acknowledgments}
The authors are grateful to Benjamin Huard, Michael Hatridge, Archana Kamal and Katrina Sliwa for valuable discussions in the preparation of this review. The authors thank Maxime Malnou for his critical reading of the manuscript. This work is supported by ARO under Grant No.W911NF-14-1-0011.

\section*{References}

\bibliographystyle{elsarticle-num}
\bibliography{library}

\end{document}